\documentclass[fleqn,twoside]{article}
\usepackage{espcrc2}

\usepackage[dvips]{graphicx}
\usepackage[figuresright]{rotating}

\newcommand{\AmS}{{\protect\the\textfont2
  A\kern-.1667em\lower.5ex\hbox{M}\kern-.125emS}}

\newcommand{\new}{\newcommand}
\new{\Mzsq}       {{\ifmmode M^2_{\mathrm{ Z}}
                    \else $M^2_{\mathrm{ Z}}$\fi}}
\new{\Mz}       {{\ifmmode M_{\mathrm{ Z}}
                    \else $M_{\mathrm{ Z}}$\fi}}
\new{\as}[1]      {{\ifmmode\alpha^{#1}_s
                    \else$\alpha^{#1}_s$\fi}}
\new{\asx}[1]      {{\ifmmode a^{#1}_s
                    \else $a^{#1}_s$\fi}}
\new{\asb}[1]     {{\ifmmode\overline{\alpha}^{#1}_s
                    \else $\overline{\alpha}^{#1}_s$\fi}}
\new{\asmz}       {{\ifmmode\alpha_s(\Mz)
                    \else $\alpha_s(\Mz)$\fi}}

\title{Measurements of $\alpha_s$  from event shapes and the four--jet rate
\thanks{Talk presented at the 31st International Conference on High Energy Physics, Amsterdam,  24-31 July 2002.}}

\author{G. Dissertori\address[ETHZ]{Institute for Particle Physics, 
        ETH Z\"urich,\\
        ETH H\"onggerberg HPK E28, CH-8093 Z\"urich,  Switzerland}%
}
       
\begin{document}

\begin{abstract}
New results from measurements of the strong coupling constant \asmz\  at LEP
are presented. In particular, a new LEP combination of results based on event--shape variables
has become available, where a new method for the estimation of the theoretical
uncertainty has been implemented. Furthermore, two other analyses are quoted,
based on power corrections and the four-jet rate. 
\vspace{1pc}
\end{abstract}

\maketitle

\section{Introduction}

The strong coupling constant \as{} is the only free parameter of the QCD Lagrangian
and many measurement have been performed at various energy scales in different processes 
in order to determine it precisely \cite{Bethke}. The measurements at LEP
have contributed significantly to this effort.

Two years after the closure of LEP still new results from QCD analyses become available 
\cite{ALEPHetal}. These results are based on data from LEP1, where the centre-of-mass energy was
around 91 GeV and the data sample consisted of several million events, practically background free, as well as from LEP2, with energies from 133 GeV up to 206 GeV. Here the data samples are much smaller, of the order of several thousand events, and the backgrounds (mainly fully hadronic 
W-pair events) can amount up to about 15\%. In addition, at LEP2 a special selection has to be
applied in order to eliminate the so-called {\em radiative return events}, where a hard photon is 
radiated in the initial state, reducing the effective centre-of-mass energy to the Z mass. This leads
to a significant loss in statistics. Therefore the \as{} measurements at LEP2 have a much
larger statistical uncertainty. However, systematic uncertainties related to non-perturbative
corrections or unknown higher order terms are reduced because of the larger energy scale.
This motivates the combination of all measurements performed by the four LEP experiments
ALEPH, DELPHI, OPAL and L3, in order to obtain a precise test of the running of the 
coupling, as well as a competitive measurement at LEP2 compared to LEP1, and finally the
best possible overall combination. Such a combination has now been performed by the
LEP QCD working group (LEPQCDWG) for measurements based on event--shape variables.

\section{Event--shape variables}

Event--shape variables are a classical tool to measure the strong coupling constant.
They are sensitive to the topology of an event, therefore to gluon radiation and thus to the
coupling strength. The most thoroughly tested variables are thrust, heavy jet mass, C--parameter,
total and wide jet broadening and the $y_3$--distribution, where $y_3$ is the resolution
parameter where a three--jet event is clustered into a two--jet event (Durham clustering algorithm). For these variables the perturbative QCD predictions
exist at next--to--leading order (NLO), $\mathcal{O}(\as{2})$. In addition, large logarithms 
$L=\ln(1/y)$ 
have been resummed at next--to--leading logarithmic (NLL) accuracy 
to all orders in \as{}, where $y$ is an event--shape variable. Only recently some missing NLL terms have been calculated for the
$y_3$--distribution \cite{Salam}. Several matching schemes have been proposed over the
years in order to combine the NLO and NLL calculations, such as the logR and R matching
schemes \cite{Catani}. The LEPQCDWG adopts so--called {\em modified matching schemes},
where the logarithmic term  is replaced by
$
 L \rightarrow \tilde L=\frac{1}{p}\ln \left[(1/y)^p -  
               (1/y_\mathrm{max})^p +1 \right] \; .
$
This ensures a physical behaviour of the prediction at the phase space boundary $y_\mathrm{max}$. The choice of the power $p$ is arbitrary, in the sense that different choices
only give different subleading contributions, whereas the NLL terms remain unchanged. The modified logR matching scheme corresponds to $p=1$, whereas $p=2$ is called the {\em second degree modification}.
\begin{figure}[thb]
 \begin{center}
  \includegraphics[width=7cm]{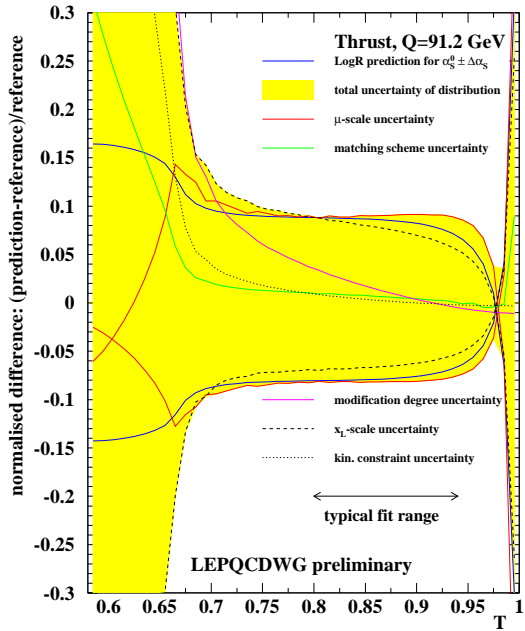}\\[-0.8cm]
  \caption{Uncertainty band for the thrust distribution. \label{unband}}
  \end{center}
  \vspace{-1cm}
\end{figure}
As mentioned earlier, the measurements at LEP1 are basically background free, whereas at 
LEP2 the W background is very difficult to suppress, in particular in the multi--jet region
of the event--shape distribution which is also sensitive to hard gluon radiation. In order to
avoid large systematic uncertainties related to the background subtraction, a careful choice
of the fit range in the event--shape distribution is needed, over which the theoretical prediction is adjusted to the data. At the other end of the spectrum (the two--jet limit) the hadronization
corrections become sizeable, and correspondingly  the fit range has to be adapted.    
These corrections are obtained from phenomenological
models implemented in Monte Carlo generators (JETSET, HERWIG, ARIADNE). The difference in the results obtained with different generators are quoted as
systematic uncertainties due to hadronization, of the order of 1-2\%.

The dominant systematic uncertainty is related to missing higher order terms in the perturbative
calculations. In order to avoid a very bad fit quality when changing the 
renormalization scale, and in particular to reduce the multitude and arbitrariness of prescriptions
used by the LEP experiments, the LEPQCDWG has proposed a new method, the so--called {\em uncertainty band method}. First a reference prediction is chosen, such as the modified logR matching scheme, together with a certain \as{} value. Then various changes are applied to the prediction, keeping the same \as{} value:
The renormalization scale $\mu$ is varied in the range $0.5 < \mu/\sqrt{s} < 2$, the modified R matching scheme is applied, the second degree modification is tested, the logarithmic terms are changed
to $L'=\ln(1/(x_L y))$ for $2/3 \le x_L \le 3/2$,  which introduces different subleading terms,
and the kinematic boundary $y_\mathrm{max}$ is varied over a range given by the difference
between NLO and parton shower predictions. The difference of the distributions thus obtained to the reference distribution defines an uncertainty band (Fig.\ \ref{unband}). In the next step again
the reference prediction is taken, but now \asmz\ is varied such that the changes in the
prediction reproduce the spread given by the uncertainty band over the fit range. The necessary
variations in \asmz\ define the theoretical uncertainty.

\begin{figure}[thb]
 \begin{center}
  \includegraphics[width=7cm]{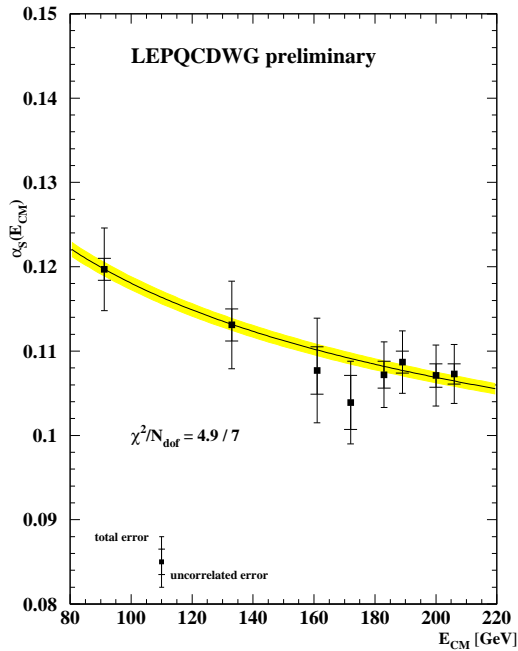}\\[-0.8cm]
  \caption{Combined measurements of \as{} from event--shape variables at different LEP energies, compared to the expected running. \label{run}}
  \end{center}
  \vspace{-1cm}
\end{figure}

The LEPQCDWG has combined all measurements based on event--shape distributions
at LEP1 and LEP2 for the six variables mentioned above. Correlations between variables, energies and experiments have been taken into account. The results are
$\asmz = 0.1197\pm0.0002_\mathrm{stat}\pm0.0008_\mathrm{exp}\pm0.0010_\mathrm{had}
\pm{\mbox{}^{0.0048}_{0.0047}}_\mathrm{theo}$ for LEP1 alone,
  $\asmz = 0.1196\pm0.0005_\mathrm{stat}\pm0.0010_\mathrm{exp}\pm0.0007_\mathrm{had}
\pm{\mbox{}^{0.0043}_{0.0044}}_\mathrm{theo}$ for LEP2 alone, and
$\asmz = 0.1198\pm0.0003_\mathrm{stat}\pm0.0009_\mathrm{exp}\pm0.0008_\mathrm{had}
\pm0.0046_\mathrm{theo}$ for all LEP energies combined (including measurements at an effective  centre-of-mass energy below \Mz, obtained by selecting events with hard initial state radiation). Excellent consistency between the
LEP1 and LEP2 measurements is found. Taking the individual measurements at the various
energy points, perfect agreement with the expected running of the strong coupling is observed
(Fig.\ \ref{run}).

In an analysis by DELPHI the energy dependence of the mean values of event--shape variables has been analysed \cite{DELPHI}, but now the hadronization corrections are replaced by a power law as proposed in \cite{DWeb}. It results in a  rather precise determination of 
$\asmz = 0.1184\pm0.0033_\mathrm{tot}$.  A more detailed discussion of power law studies can be found in \cite{Movilla}.

\section{Four--jet rate}

Recently, the ALEPH collaboration has published a new determination of \as{} \cite{Aleph4jet},
based on a measurement of the four-jet rate (Durham algorithm) at LEP1. The difference to 
the event--shape variables is that here the perturbative expansion starts at $\mathcal{O}(\as{2})$.
The full NLO corrections are known, as well as the resummation of NLL terms \cite{Zoltanetal}.
The quoted result is extremely precise, namely
$\asmz = 0.1170\pm0.0001_\mathrm{stat}\pm0.0009_\mathrm{exp}\pm0.0003_\mathrm{had}
\pm0.0008_\mathrm{theo}$. This is mainly due to the
larger sensitivity, since the leading term is of  $\mathcal{O}(\as{2})$. It is observed that although the $\chi^2$ of the fit changes substantially when varying the renormalization scale over a wide range,
and also  the location of the sharp minimum in this $\chi^2$ varies after  changes in the fitted prediction, the actually fitted \asmz\ value remains remarkably stable.

\section{Summary}

New combined results for measurements of \asmz\ at LEP have been presented, based on event--shape distributions and a new prescription for the theoretical uncertainty. Furthermore, 
it has turned out that the measurement of the four-jet rate at  LEP allows for a very precise
determination of \asmz.

\section{Acknowledgements}

I would like to thank the members of the LEPQCDWG for providing me with the relevant
numbers and plots.


\begin{thebibliography}{9}
\bibitem{Bethke}  S.\ Bethke, {\em J.\ Phys.\/} {\bf G26} (2000) R27. 
\bibitem{ALEPHetal} ALEPH Collab., Note 2002-012 CONF 2002-002, ICHEP02 abs296; 
L3 Collab., {\em Phys.\ Lett.}~ {\bf B536} (2002) 217; DELPHI Collab., Note 2002-050-CONF-584, ICHEP02 abs229; OPAL Collab., Note PN512, ICHEP02 abs368;  Note PN473, ICHEP02 abs370.
\bibitem{Salam} A.\ Banfi {\em et~al.}, {\em JHEP}~ {\bf 0201} (2002) 018.
\bibitem{Catani} S.\ Catani {\em et~al.}, {\em Nucl.\ Phys.\/},~{\bf B407} (1993) 3.
\bibitem{DELPHI} DELPHI Collab.,  Note 2002-050-CONF-584, ICHEP02 abs229.
\bibitem{DWeb} Yu.L.\ Dokshitzer and B.R.\ Webber, {\em Phys.\ Lett.\/}~{\bf B352} (1995) 451.
\bibitem{Movilla} P.A.\ Movilla Fernandez, these proceedings.
\bibitem{Aleph4jet} ALEPH Collab., CERN-EP-2002-029, accepted for publication in {\em Eur.\ Phys.\  J. \ C}.
\bibitem{Zoltanetal} Z.\ Nagy and Z.\ Tr\'{o}cs\'{a}nyi,  {\em Phys.\ Rev.\ Lett.\/}~{\bf 79} (1997) 3604; {\em Phys.\ Rev.\/}~{\bf D59} (1999) 014020;\\
 L.~J.\ Dixon and A.\ Signer, {\em Phys.\ Rev.\ Lett.\/}~{\bf 78} (1997) 811.
\end{thebibliography}
 \end{document}